\documentstyle[aps,prb,multicol,epsf]{revtex} 
\begin{document}   
\draft 
\title{Weak localization in disordered systems at the ballistic 
limit}     
                   
\author{Assaf Ater and Oded Agam}  

\address{The Racah Institute of Physics, The Hebrew University, Jerusalem 
91904 Israel} 
\maketitle 
\begin{abstract}
The weak localization (WL) contribution to the   
two-level correlation function, $R(\omega)$, is calculated for 
two-dimensional disordered conductors. Our analysis extends to the 
nondiffusive (ballistic) regime, 
where the elastic mean path is of order of the size of the system. 
In this regime the structure factor, $S(t)$, (the Fourier transform of 
$R(\omega)$) 
exhibits a singular behavior consisting of dips superimposed on  
a smooth positive background. The strongest dips appear at periods of 
the periodic orbits of the underlying clean system. Somewhat weaker 
singularities 
appear at times which are sums of periods of two such orbits.
The results elucidate various aspects of the weak localization
physics of ballistic chaotic systems.
\end{abstract}      
\begin{multicols}{2} 

\section{Introduction}
Interference effects, arising from the interplay of phases 
accumulated along different paths, are particularly interesting 
in ballistic chaotic systems. It is due to the
hierarchy of importance among the classical trajectories in these 
systems: Long trajectories exhibit a universal statistical 
behavior, while short ones constitute the dynamical 
fingerprints of the system. 
The stable (and therefore usually also the shorter)
the orbit is the stronger is its signature. 
This signature appears
both in the wave functions (the scar phenomenon\cite{Heller84}) 
as well as in the statistical properties of the 
energy spectrum of the system\cite{Berry85}.
The purpose of this paper is to study the fingerprints of the 
classical periodic orbits on the nature of interference in
chaotic systems.

Our best understanding of quantum interference is in 
disordered systems. In these systems 
interference may lead to localization of the particle 
in space\cite{Anderson58}. If, however, the disorder is too weak to 
localize the particle, interference manifest itself 
as an increase in the return probability compared to the classical value.   
This effect, known as weak localization (WL), has been observed by   
measuring the magnetoresistance of metallic films\cite{Bergmann}.    
   
Recent advances in nanostructure technology\cite{advances},   
opened the possibility of manufacturing clean mesoscopic systems --    
systems in which the elastic mean free path, $l$, is of order of the   
size of the system, $L$. It is natural to ask what is    
the analogue of WL in such ballistic systems?  
    
Very little is known about this issue,
mainly because of the failure of periodic orbit theory    
to provide a simple systematic procedure for calculating 
interference (i.e.~WL) corrections\cite{Whitney99}. 
This failure has been one of the main
motivations for constructing the supersymmetric nonlinear $\sigma$-model
of ballistic systems\cite{NLSM}. The hope was that this model will 
produce a WL expansion for ballistic systems analogous to that
of disordered systems. However, it turned out that
WL crucially depends on the regularization of the field integral,
and only specific cases could be worked out\cite{Aleiner}. 
These are the cases where the dynamics   
is still diffusive or dictated by random matrix 
theory (RMT)\cite{Mehta91}.   
    
Usually one would choose to study the WL signature in transport properties,
because they are naturally related to the experimental data. However,
this choice will be inappropriate for our purpose for the following reason:
WL (similar to localization) takes place on a certain manifold in phase space. 
For example, in disordered systems this manifold is the  real space, while 
in a circular billiard with rough boundaries localization
occurs in the angular momentum space\cite{roughb}. In general 
chaotic systems there is no preferred basis, therefore,  
WL may appear on a complicated manifold in the phase space\cite{Shepelyansky86}. 
Yet, transport measurements dictate a preferred basis,
and may totally miss the WL physics 
we seek to describe.

Nevertheless, interference effects manifest themselves also in the 
spectral properties of chaotic systems, which are  
independent of the choice of basis. Therefore, in this work we shall focus our   
attention on the WL contribution to the 
simplest nontrivial spectral quantity -- the two-level 
correlation function:  
\begin{equation}   
R(\omega)= \Delta^2 \left\langle    
\rho(\epsilon+ \hbar \omega)\rho(\epsilon) \right\rangle -1. \label{Rw}   
\end{equation}   
Here $\rho(\epsilon)= \sum_\alpha\delta(\epsilon -\epsilon_\alpha)$ is   
the density of states, $\Delta= 1/\langle \rho \rangle$ 
is the mean spacing between   
neighboring energy levels, $\epsilon_\alpha$, and the averaging,    
$\langle \cdots \rangle$, is over the disorder configurations 
or the energy $\epsilon$.   

To state our problem in this context, consider the density of 
states of quantum system with Hamiltonian having a classical
chaotic counterpart. Gutzwiller's trace formula \cite{Gutzwiller} 
expresses the density of states, in the semiclassical limit, as a 
sum over the classical periodic orbits of  
the system:  
\begin{eqnarray}  
\rho(\epsilon) \simeq \frac{1}{\Delta} + \sum_{p.o.}  
A_{p} e^{\frac{i}{\hbar}s_{p}(\epsilon )}, \label{Gutz}  
\end{eqnarray}  
where $s_p(\epsilon )$ 
is the action of the $p$-th periodic orbit,  
and $A_p$ is the corresponding amplitude depending on the 
stability of the orbit and its period \cite{Gutzwiller}. 

The traditional way of calculating correlators such as (\ref{Rw}),
within periodic orbit theory, is to use the so called  
diagonal approximation \cite{Hannay84,Berry85}.  
In this approximation one 
{\narrowtext    
\begin{figure}      
  \begin{center}     
\leavevmode     
        \epsfxsize=5.33cm        
         \epsfbox{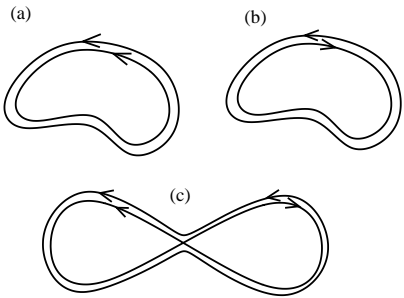}      
\end{center}     
   
	\caption{(a+b) An illustration of pairs of orbits contributing 
to the diagonal approximation: (a) an orbit with itself,  
(b) an orbit with its time reversed counterpart.
(c)  The eight-shaped periodic orbits associated with the WL
contribution to $R(\omega)$ in diffusive systems.
}    
\label{fig:diaorbits}       
\end{figure}        
} 
\noindent
replaces a double   
sum over periodic orbits (such as that obtained when substituting
(\ref{Gutz}) in (\ref{Rw}))  by a single sum: 
\begin{eqnarray}  
 \left\langle \sum_{p p'} A_pA^*_{p'} e^{\frac{i}{\hbar}\left[ s_p(\epsilon + 
 \hbar\omega ) - s_{p'}(\epsilon)\right]}  
\right\rangle  \to \frac{2}{\beta}  
\sum_p |A_p|^2 e^{i\omega \tau_p }, \label{diagonalaprox1}   
\end{eqnarray} 
where $\tau_p= \partial s_p(\epsilon)/\partial 
\epsilon $ is the period of the $p$-th orbit. 
The rational behind the diagonal approximation is that 
the coherent contributions, at $\omega=0$, come
from pairs of orbits $(p,p')$ 
having precisely the same action. Thus one should pair orbits  
with themselves, $p=p'$ (Fig.~\ref{fig:diaorbits}a), as well as with other 
orbits related by symmetries such as time reversal symmetry (Fig.~\ref{fig:diaorbits}b).
In the absence of other spatial symmetries,  $\beta$ in the above 
formula is one for systems with time reversal 
symmetry, and two for systems which do not have this symmetry.

The problem of WL in the context of the two-level 
correlation function can be formulated as:  
{\em How can one improve on the diagonal approximation to include 
interference effects systematically?} 

In seeking for the solution of this 
problem it is natural to inquire about the situation in disordered systems 
where the systematic interference corrections to the diagonal approximation 
is the ``weak localization'' expansion. 
The diagrammatic picture of the WL correction to
$R(\omega)$ suggests that the WL contribution is associated with 
pairs of periodic orbits crossing themselves at some point in space
as shown in Fig.~\ref{fig:diaorbits}c\cite{Smith98}. 
Thus along one loop the two orbits propagate
in the same direction, while along the other loop they are in opposite directions. 
However, such orbits exist only in the presence of a 
non-classical scattering potential, and do not have a
direct analog in the periodic orbit theory.  
     
Facing this difficulty, in this work, we  
study WL using disorder diagrammatics but far from the   
diffusive regime, i.e. when the elastic mean free path is of 
order of the size of the system. In this case, the disorder 
is sufficiently weak,  and traces of the short periodic orbits
of the underlying clean system are still significant.

We, thus, consider a system consisting of a particle 
confined to move on a two  dimensional torus, in the landscape of a 
random potential, see Fig.~\ref{fig:system}. The Hamiltonian 
{\narrowtext    
\begin{figure}      
  \begin{center}     
\leavevmode     
        \epsfxsize=5.33cm        
         \epsfbox{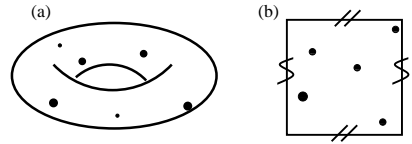}      
\end{center}     
   
	\caption{(a) An illustration of the model used in this paper 
for calculating the WL effects in the ballistic limit.
The system consists of a non-interacting electron gas 
on a torus with white noise potential. This potential is sufficiently weak, 
such that the elastic mean free path is of order 
of the size of the system. (b) An equivalent representation of
the system as a square with periodic boundary conditions.} 
\label{fig:system}  
\end{figure}        
} 

\noindent
of the system is 
\begin{equation}  
\label{eq:hamiltonian}  
H=\frac{{\bf p}^{2}}{2m}+V({\bf r}), 
\end{equation}
where ${\bf p}$ is the momentum of the particle, $m$ is its mass,  and 
$V({\bf r})$ is a Gaussian random potential defined by
\begin{eqnarray}
\label{eq:potential} 
\langle V({\bf r}) \rangle &=&0, ~~~~\mbox{and}~~~~~ 
\langle V({\bf r})V({\bf r}') \rangle =
\frac{\hbar}{2\pi \nu \tau}\delta({\bf r}-{\bf r}'). 
\end{eqnarray} 
Here, $\nu=1/\Delta L^2$ is the averaged density 
of states per unit area, and $\tau$ is the elastic mean 
free time for scattering 
on the potential.
This system has been considered earlier by Altland and 
Gefen~\cite{Altland93} and by  Agam and Fishman~\cite{Agam96d},
but only in the framework of the diagonal approximation.  

In analyzing the results of the above model, it will be
convenient to use the spectral structure factor defined as
\begin{equation} 
 S(t)=\frac{\hbar}{\Delta}\int_{-\infty}^{\infty}d\omega  
R(\omega)e^{-i \omega t}.\label{eq:sdef}
\end{equation}
Using $S(t)$ one can relate the 
quantum spectral properties of the system to the behavior 
of its classical analog. In particular, $S(t)$ form 
a connection to the periodic orbits of the system: 
Substituting (\ref{diagonalaprox1}) in 
(\ref{eq:sdef}) one sees that, within the diagonal approximation,
the structure factor takes the form of 
a sum over peaks located at times which equal to the periods 
of the classical periodic orbits: 
\begin{equation}
S(t)\simeq\frac{2h\Delta}{\beta}\sum_{p}|A_{p}|^{2}\delta(t-\tau_{p}).
\end{equation}
It has been noticed by Argaman et al.\cite{argaman} that
the right hand side of the above equation can be also interpreted
as $|t| p(t)$, where $p(t)$ is the classical return probability 
at time $t$. The notion of return probability has been further 
developed by Chalker et al.~to obtain a more accurate 
description of the structure factor for diffusive electrons\cite{chalker}.

A disorder potential usually erases the $\delta$-singularities of $S(t)$ 
associated with the classical orbits of the clean system. But, if it is 
sufficiently weak, it will leave traces of the them. 
Indeed, $S(t)$ calculated, in the diagonal approximation, 
for weak disorder, shows a series of peaks~\cite{Agam96d}
 (see inset of Fig.~3). 
The locations of these peaks
{\narrowtext   
\begin{figure} 
    
  \begin{center} 
  \vspace{-3.0cm}     
\leavevmode    
       \epsfxsize=7.33cm        
         \epsfbox{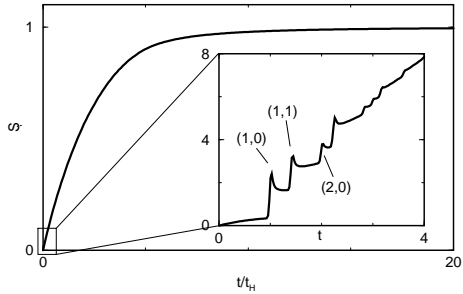}      
\end{center}     
   
	\caption{The structure factor of chaotic systems with time 
reversal symmetry. The solid line represents the results of 
random matrix theory. Magnified is the regime where 
perturbation theory applies and nonuniversal corrections, which are
the main focus of this paper, are important. 
Here we depict only the results of the contribution of the
``diagonal approximation''. The peaks, indicated by pairs of 
winding numbers, are  the signatures of the periodic orbits of
the clean system (see Fig.~4). Both, the Fermi velocity and the system size, 
$L$, are set to unity. The elastic mean  free path, in these units, is 1/2.
$t_H=2 \pi \hbar/\Delta$ is the Heisenberg time.}
\label{fig:rmtformfactor} 
\end{figure} 
} 
 
\noindent 
along the time axis are precisely the periods of the orbits of the clean system. 
(These orbits are defined by pairs of winding numbers which count the
times the trajectory winds around the torus 
in each direction, see Fig.~\ref{fig:w.n}.)

In view of the behavior shown in Fig.~3 and the results of disorder
diagrammatics, one may naively
speculate that the WL contribution to the structure factor 
adds up in a similar way. Namely it 
consists of a series of singularities located at periods of 
the eight-shaped orbits illustrated in Fig.~\ref{fig:diaorbits}c.
One may also expect this contribution to be positive, as  
in diffusive systems, since it should reflect an increase 
in the return probability compared to the classical 
value (i.e.~the diagonal approximation). 

However, as we show here, this picture is inaccurate. 
Indeed, in the ballistic regime,  some singularities do 
appear at times which can be interpreted as
periods of eight-shaped orbits (Fig.~1c). But
these contributions are rather weak. A large {\it negative} 
contribution comes from the original periodic orbits. It 
is superimposed on a smooth positive background which is 
not related to properties of the clean system. 
At certain cases the WL contribution 
to the structure factor can even become altogether negative. 
Thus, in ballistic systems, it does not have, necessarily, 
a definite sign. 

To make the paper self-contained, we organized it
as follows: In the next section we prepare
the mathematical background for our derivation by reviewing the standard 
results of disorder diagrammatics in the diffusive limit. This
way we set the basis for extending the diagrammatic approach 
to the ballistic regime. In section III we  
{\narrowtext    
\begin{figure}      
  \begin{center}     
\leavevmode     
        \epsfxsize=7.33cm        
         \epsfbox{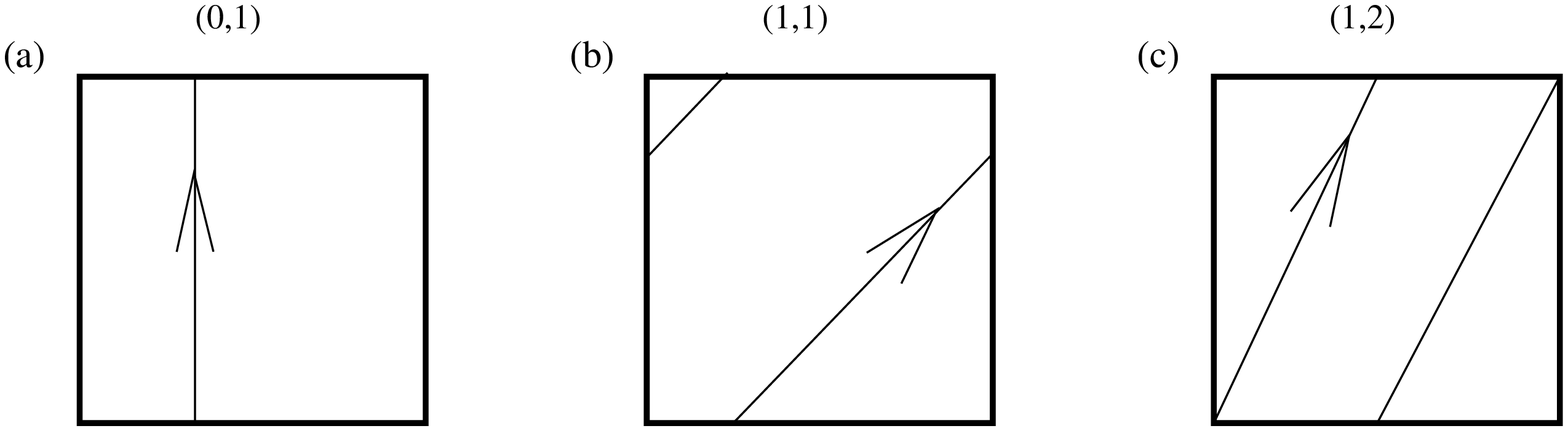}      
\end{center}     
   
	\caption{Periodic orbits of a particle moving on a  
tours are defined by pairs of integer numbers $(n_x,n_y)$. 
These ``winding numbers'' count the number
of times the trajectory winds around the torus in the $x$ and in 
the $y$  directions, respectively. Some particular examples are:
(a) the orbit (0,1) with length $L$,  
(b) the orbit (1,1) of length $2^{\frac{1}{2}} L $, and 
(c) the orbit (1,2) of length $5^{\frac{1}{2}} L$.} 
      \label{fig:w.n} 
\end{figure} 
} 

\noindent
derive our central formulae for the WL contribution to $R(\omega)$, and the
structure factor, $S(t)$. In
section IV, we analyze these results and derive 
an asymptotic expression for $S(t)$. Finally, we summarize and
present our conclusions in Section V.

\section{Background} 
The purpose of this section is to lay the 
technical background, and set the nomenclature for the analysis which   
will be carried out in the forthcoming sections. We shall 
review the main ideas of disorder diagrammatic technique for diffusive 
systems\cite{Abrikosov}, present the basic building blocks, 
discuss the approximations involved, and the    
limits of applicability. Finally, we summarize the 
results for the  WL contribution to $R(\omega)$ within RMT
framework, and for diffusive systems.
These results will form a reference point for 
the analysis of $R(\omega)$ in the ballistic limit, which 
will be carried out in the next section.   
   
The disorder diagrammatic approach for Hamiltonians of the    
type (\ref{eq:hamiltonian}) is an efficient way of    
constructing the perturbation expansion, in the weak potential 
$V({\bf r})$,    
for quantities averaged over the disorder configurations.    
Examples of such quantities are $n$-point spectral correlation   
functions, the magnetic susceptibility,    
and various properties of the conductance of disordered metals.   
   
This diagrammatic approach is a semiclassical approximation in which 
the ratio of the particle wave length, $\lambda_{F}$,   
to the elastic mean free path, $l$, is assumed to be small.    
Therefore, it takes the formal form of an asymptotic series in powers    
of $1/k_Fl$, where $k_F=2 \pi/\lambda_{F}$ is the Fermi wavenumber. 
Yet, usually there will be also
non-perturbative contributions,  which are important when trying    
to resolve features on the scale of the mean level spacing, 
$\Delta$, or over time scales longer than the   
Heisenberg time $t_H= 2 \pi \hbar/\Delta$. Therefore, the 
applicability range of disorder diagrammatic is also limited 
to times smaller than the  Heisenberg time, and energies larger 
than the mean level spacing.

As a first example, consider the   
average of the retarded Green function:

{\narrowtext    
\begin{figure}      
  \begin{center}     
\leavevmode     
        \epsfxsize=8.33cm        
         \epsfbox{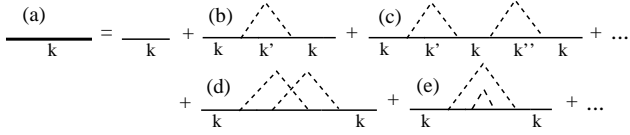}      
\end{center}     
   
	\caption{Example of diagrams contributing to the average
Green function, $G_\epsilon^R({\bf k})$ (represented by the bold line). 
Thin lines represent the free Green function (i.e.~in the absence of disorder),
and dashed lines represents impurity scatterers.}     
	\label{fig:dressed}   
\end{figure}   
} 
\noindent
\begin{eqnarray}   
G^{R}_\epsilon({\bf k}) =\left\langle    
\frac{1}{\epsilon+i\eta-\frac{\hbar^2{\bf k}^2}{2 m} -
 V({\bf r})} \right\rangle. \nonumber
\end{eqnarray}   
Here $\langle \cdots \rangle$ denotes an averaging over the    
configurations of the disordered potential, 
$\eta$ is an infinitesimal positive number,   
and ${\bf p}$ is the particle momentum. Expanding 
the Green function in powers of $V({\bf r})$, and changing 
representation to momentum space yields   
\begin{eqnarray}   
 G^{R}_\epsilon({\bf k}) = G^{R}_{0}(\epsilon,{\bf k})   
+ G^{R}_{0}(\epsilon,{\bf k})   
\left\langle V_0 \right\rangle G^{R}_{0}(\epsilon,{\bf k}) + ~~~~~ 
\nonumber\\   
+\sum_{{\bf k'}}G^{R}_{0}(\epsilon,{\bf k})G^{R}_{0}(\epsilon,{\bf k'})
G^{R}_{0} (\epsilon,{\bf k})
\left\langle V_{\bf k-k'}   
V_{\bf k'-k}\right\rangle+...   \nonumber
\end{eqnarray}   
where $V_{\bf q}=\frac{1}{L^{2}}\int d^{2}r 
e^{\frac{i}{\hbar}{\bf q \cdot r}}V({\bf r})$ is the Fourier    
transform of the potential (\ref{eq:potential}), 
and $G^R_0(\epsilon,{\bf k})   
= 1/(\epsilon+i\eta-\frac{\hbar^2 k^2}{2 m})$ is the free Green function.  
Terms containing an odd number of $V$-s vanish upon averaging,
while those having an even number are calculated  
by Wick's theorem (since the potential $V$ is Gaussian). Thus 
the average is equal to the product of averages of all possible   
pairs, such as $\langle V_{\bf q} V_{-{\bf q}} \rangle$. The various terms
of this expansion can be represented  diagrammatically as shown 
in Fig.~\ref{fig:dressed}.   
  
A partial summation of the infinite series of the diagrams in 
Fig.~\ref{fig:dressed}, is achieved using Dyson's equation, and
summation over the irreducible diagrams (those which cannot be    
separated into two disconnected diagrams by cutting one   
internal propagator line, e.g. (b) (d) and (e) in   
Fig.~\ref{fig:dressed}). 
Thus the averaged Green function satisfies the relation    
\begin{equation}   
G^{R}_\epsilon({\bf k})=G^{R}_{0}
(\epsilon,{\bf k})+G^{R}_\epsilon({\bf k})\Sigma   
G^{R}_{0}(\epsilon,{\bf k}),\label{eq:dysongreen}   
\end{equation}   
where $\Sigma$ is the self energy given by the sum over all irreducible    
diagrams, see Fig.~\ref{fig:dresseddyson}.   
To the leading order in $1/k_{F}l$,  $\Sigma$ is 
the contribution of the first diagram in 
Fig.~\ref{fig:dresseddyson}b. Thus  
\begin{eqnarray}   
\Sigma&\simeq&\sum_{{\bf q}}\left\langle V_{\bf q}
V_{-{\bf q}}\right\rangle   
G_{0}(\epsilon,{\bf k}+{\bf q})\nonumber \\   
&=&\frac{\hbar\Delta}{2\pi\tau}\left[P.V.\left(\int   
d\xi \frac{\rho(\xi)}{\epsilon-\xi+i\eta}\right)  
-\frac{i\pi}{\Delta}\right] \nonumber,   
\end{eqnarray}   
where $P.V.$  denotes the principle value of the integral.
The real part of $\Sigma$ can be absorbed into the definition   
of the reference energy $\epsilon$, thus the solution of Dyson's   
equation (\ref{eq:dysongreen}) yields

{\narrowtext    
\begin{figure}      
  \begin{center}     
\leavevmode     
        \epsfxsize=8.33cm        
         \epsfbox{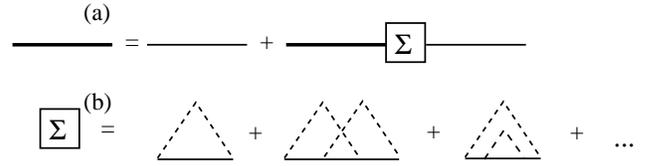}      
\end{center}     
   
	\caption{(a) The diagrammatic representation of Dyson's equation 
for the average Green function (\ref{eq:dysongreen}). The bold 
and thin lines represents the dressed 
and the bare Green functions, respectively. (b) 
The self energy, $\Sigma$, given as a sum
of irreducible diagrams. }   
	\label{fig:dresseddyson}    
\end{figure}
}
\noindent
\begin{eqnarray}   
G^{R}_\epsilon({\bf k})=\frac{1}{\epsilon\!-\!\epsilon({\bf k})\!+\!
\frac{i\hbar}{2\tau}},    \nonumber  
\end{eqnarray}   
where $\epsilon({\bf k})=(\hbar k)^{2}/2m$. 
Similarly, the average of the 
advanced Green function is given by 
$G^{A}_\epsilon({\bf k})= (G^{R}_\epsilon({\bf k}))^{*}$.   
 
Consider, next, the probability of    
a particle to arrive to ${\bf r}'$ in time $t$, when its  
initial state, $| {\bf r}; \epsilon_F \rangle$, 
is a wave packet localized near ${\bf r}$. We assume that this
wave packet is composed of eigenstates centered at the Fermi energy, 
$\epsilon_F$, 
and ranging over an energy band of width $\hbar/\tau$. 
In the semiclassical limit  
$\hbar/\tau \ll \epsilon_F$, these conditions imply that  
the particle velocity, $v_F$, is well defined, and the   
wave packet  width is of order of the 
elastic mean free path, $l=v_F \tau$.  
The probability density for finding the particle at point ${\bf r}'$   
after time $t$ is given by   
$P({\bf r}', {\bf r};t)=L^{2}|U({\bf r}', {\bf r};t)|^2$  
where $U({\bf r}', {\bf r};t) = \langle {\bf r}'| e^{-\frac{i}{\hbar} 
H t} |{\bf r}; \epsilon_F\rangle$ is the propagator of the system.   
Using the convolution   
theorem, one obtains 
\begin{equation}  
P({\bf r}', {\bf r};t) =\hbar\int d \omega    
e^{-i\omega t} \int d \epsilon   
\tilde{\cal D} ({\bf r}', {\bf r};\omega) \label{prob}  
\end{equation}  
where  
\begin{equation}  
\tilde{\cal D}({\bf r}', {\bf r};\omega) =L^{2} \left\langle  
G^{R}({\bf r}',{\bf r};   
\epsilon_F+ \hbar\omega)  
G^{A}({\bf r},{\bf r}'; \epsilon_F) \right\rangle, \label{spacediffuson} 
\end{equation} 
and $G^{R}({\bf r}',{\bf r}; \epsilon)$ and $G^{A}({\bf r}',{\bf r}; \epsilon)$
are the exact Green functions of the system for particular realization of 
the disordered potential. Notice that under our assumptions, $\tilde{\cal D} 
({\bf r}',{\bf r};\omega)$  
weakly depends on $\epsilon$, therefore the integration over   
$\epsilon$ results in a factor of $\hbar/\tau$.

The diagrammatic expansion of $\tilde{\cal D}({\bf r}',{\bf r};\omega)$   
proceeds along the same lines described above. 
It is convenient to perform the calculation in  
Fourier space, i.e. for 
\begin{eqnarray}  
{\cal D}({\bf q},\omega) =
\left(\frac{\hbar}{2\pi\nu\tau}\right)^{2}\frac{1}{L^{2}} 
\int d{\bf r}e^{i {\bf q}\cdot {\bf r}}  
\tilde{\cal D}({\bf r}',{\bf r}'+{\bf r} ;\omega).  
\end{eqnarray}  
The leading contribution to ${\cal D}({\bf q};\omega)$, 
known as the {\it diffuson}, is given by the set of  
diagrams shown in Fig.~\ref{fig:exactdigrams}a. 
The Dyson equation summing this set of diagrams yields
\begin{eqnarray}    
{\cal D}(\bf q,\omega)&=&
\frac{\hbar}{2\pi\nu\tau}\frac{1}{1-\Pi(\omega,\bf q)}, \nonumber   
\end{eqnarray}    
 where     
{\narrowtext    
\begin{figure}      
  \begin{center}     
\leavevmode     
        \epsfxsize=8.33cm        
         \epsfbox{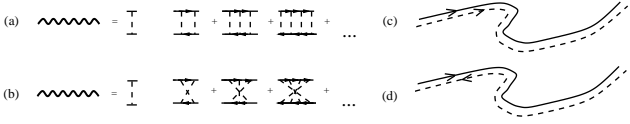}      
\end{center}     
  \caption{The diagrams of
the  diffuson (a) and  the Cooperon (b), 
and their interpretations as the contribution of 
pairs of classical orbits associated with the retarded and the advanced 
Green functions: (c) An orbit with itself (diffuson) 
and (d) an orbit with its time reversed counterpart (Cooperon).
}\label{fig:exactdigrams}    
\end{figure}      
} 
\begin{eqnarray}    
\Pi(\omega,{\bf q})&=&\frac{\hbar}{2 \pi \nu \tau}  
\int\frac{d^{2}k}{(2\pi)^{2}}G^{R}_{\epsilon+\hbar\omega}({\bf k}+{\bf q})G^{A  
}_\epsilon({\bf k}) \nonumber\\    
&\!\!\!\!\!\!\!\!\!\!\!\!\!\!\!=&\!\!\!\!\!\!\!\! 
\frac{\hbar}{2 \pi \tau}\int \frac{d\xi d\theta}{2\pi} \frac{1}{\xi +  
\hbar\omega \!-\!\hbar v_{F}q\cos\theta\!+\!\frac{i\hbar}{2\tau}}\frac{1}
{\xi\!-\!\frac{i\hbar}{2\tau}}.  
\label{eq:Pi}    
\end{eqnarray}    
To obtain the second line of the above formula,
we have expanded $({\bf k}+{\bf q})^{2}$ to linear order in   
${\bf q}$, and approximated $2 {\bf k \cdot q}$ by $2k_{F}q\cos\theta$, where  
$k_F$ is Fermi wave number, and $\theta$ is the angle between the vectors  
${\bf k}$ and ${\bf q}$. This approximation is valid when $q \ll k_F$.  
 
In the diffusive limit,  additional approximations can be made. 
Namely, one may use the small parameters 
\begin{equation}
ql \ll 1 ~~~\mbox{and}  ~~~\omega\tau \ll 1, \label{diflim}
\end{equation}
to expand
$\Pi(\omega,{\bf q})$ in $\omega \tau $, and $ql$.
The result takes  the form  $\Pi(\omega,{\bf q}) \simeq  1+
 i \omega \tau - D q^2 \tau$,  
where $D=l^2/2\tau$ is the  diffusion constant, thus
\begin{eqnarray}    
{\cal D}({\bf q},\omega)
=\frac{\hbar}{2\pi\nu\tau}~\frac{1}{-i\omega\tau+ Dq^{2}\tau}. 
\label{diffuson}   
\end{eqnarray}    
This formula shows that the diffuson is the kernel of the diffusion 
equation: $ \partial n/\partial t = D \nabla^2 n$, 
where $n({\bf r})$ is the density of particles in real space. 
The diffuson is, therefore, the classical mode of a disordered 
system in the limit of long time ($\omega \tau \ll 1$) and 
large spatial scale ( $ql \ll 1$).
  
It is instructive to relate the
diffuson to classical orbits\cite{Agam00}. 
For this purpose we turn to calculate $D({\bf r}, \omega)$ using
the van-Vleck approximation for the Green functions.  
A comment is now in order. The use of the van-Vleck propagator for a system 
with a white noise potential is unjustified, since the scattering 
is not semiclassical. Therefore, here,  we assume 
the disorder potential to be in the form of randomly located 
hard scatterers of size larger than the particle wave length.  
This potential is semiclassical, and produces diffusion on large 
scales of time and space.
 
The van-Vleck's formula for the Green function, 
$G^{R,A}({\bf r'},{\bf r}; \epsilon)$ is expressed as  
a sum over the classical trajectories\cite{Berry72}  
from ${\bf r}$ to ${\bf r'}$ with energy $\epsilon$: 
\begin{equation}  
\begin{array}{l}   
 G^R({\bf r'},{\bf r}; \epsilon)  \simeq  \frac{1}{\sqrt{2 \pi\hbar}} 
\sum_{\mu} B_\mu e^{ \frac{i}{\hbar}    
s_\mu({\bf r'},{\bf r};\epsilon)}, \\  
G^A({\bf r},{\bf r'}; \epsilon)  \simeq  \frac{1}{\sqrt{2 \pi\hbar }} 
\sum_{\mu} B_\mu^* e^{ -\frac{i}{\hbar}    
s_\mu({\bf r'},{\bf r};\epsilon)}.\label{eq:classicalgreen}  
\end{array}                             
\end{equation}    
Here $s_\mu({\bf r'},{\bf r};\epsilon)$ is the classical action  
of the $\mu$-th trajectory, while $B_\mu$ is the corresponding amplitude   
which can
 {\narrowtext    
\begin{figure}      
  \begin{center}     
\leavevmode     
        \epsfxsize=6.33cm        
         \epsfbox{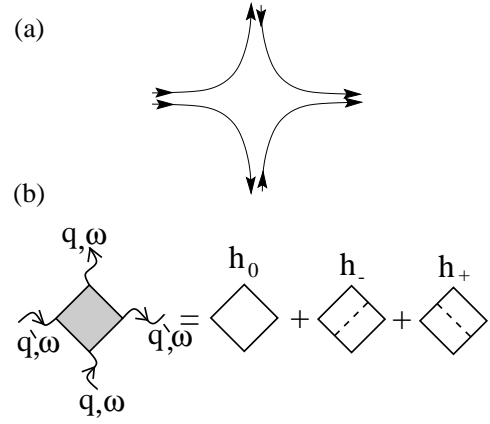}      
\end{center}     
   
	\caption{The Hikami box associated with the interaction 
between diffusons and Cooperons: (a) Its pictorial view in terms of
``classical'' trajectories. (b) Its diagrammatic expansion.}  
\label{fig:HBox}  
\end{figure}    
} 
 
\noindent
 be interpreted as the square root of the classical probability
to arrive to ${\bf r}'$,
after time $t$, starting from {\bf r}. 
Substituting (\ref{eq:classicalgreen}) in 
(\ref{spacediffuson}), yields $\tilde{\cal D}({\bf r}',{\bf r};\omega)$ 
as a double sum over the classical trajectories from ${\bf r}$
 to ${\bf r}'$. Approximating the average of this double sum by the 
diagonal part, and substituting the result in (\ref{prob}) we obtain 
\begin{eqnarray}  
P({\bf r}',{\bf r},t)\simeq L^{2}\sum_{\mu}  
|B_{\mu}|^{2}\delta(t-\tau_{\mu}), \nonumber  
\end{eqnarray}  
where $\tau_\mu$ is the time which takes for the particle to travel from  
${\bf r}$ to ${\bf r}'$ along the $\mu$-th trajectory.
Using classical sum rules, one can sum over the classical 
trajectories\cite{Agam00}. The result for 
diffusive systems is that of the diagrammatic
calculation. This implies that set of diagrams associated with 
the diffuson is equivalent to the diagonal approximation of 
pairs of orbits as shown in Fig.~\ref{fig:exactdigrams}c.  
 
In systems with time reversal symmetry there is an additional  
classical mode called Cooperon. It comes from the  
infinite sum over the maximally crossed diagrams shown in  
Fig.~\ref{fig:exactdigrams}b. These diagrams   
are obtained by reversing the direction of the momentum in one of the 
Green function lines. The classical picture of the Cooperon is, 
therefore, that of an orbit paired with its 
time reversed counterpart as shown in  
Fig.~\ref{fig:exactdigrams}d. It can be easily checked that the Cooperon 
has precisely the same analytical form of the diffuson.   
 
The issue of WL, in the language of diagrammatics,
is the interaction between diffuson and Coopron modes. 
Pictorially, this interaction is the switching between the
directions of the momenta of two trajectories, as  
shown in Figs.~1c and 8a. The diagrammatic entity accounting for this 
switching is the Hikami box\cite{Hikami81}, see 
Fig.~\ref{fig:HBox}b.  It is a function of the incoming 
and outgoing momenta and frequencies of the diffuson and the Cooperon. 
For the particular choice of momenta and frequencies shown in
Fig.~\ref{fig:HBox}b one has 
$h({\bf q},{\bf q}',\omega) =h_{0}+h_{-}+h_{+}$,  where  
\begin{eqnarray}  
h_{0} &=&\sum_{\bf k}    
 G^{R}_{\epsilon + \hbar \omega}({\bf k})G^{R}_{\epsilon+ \hbar \omega} 
({\bf k}\!-\!{\bf q}\!-\!{\bf q}')    
G^{A}_\epsilon({\bf k}\!-\!{\bf q})G^{A}_\epsilon ({\bf k}\!-\!{\bf q}'),  
\nonumber \\  
h_{-} &=&
\frac{\hbar\Delta}{2\pi\tau} h_{\frac{1}{2}}({\bf q},-{\bf q}';\epsilon, 
\epsilon+ \hbar \omega) 
 h_{\frac{1}{2}}(-{\bf q},{\bf q}';\epsilon, \epsilon+ \hbar \omega) 
\label{Hbox}
 \\  
h_{+} &=& \frac{\hbar\Delta}{2\pi\tau} h_{\frac{1}{2}}^*
({\bf q},{\bf q}'; \epsilon+ \hbar \omega, \epsilon)  
h_{\frac{1}{2}}^*(-{\bf q}',-{\bf q}; \epsilon+ \hbar \omega, \epsilon), 
\nonumber 
\end{eqnarray}
while
\begin{eqnarray}
h_{\frac{1}{2}}({\bf q},{\bf q}'; \epsilon, \epsilon') = 
\sum_{\bf k} G^{A}_{\epsilon}({\bf k}) 
G^{R}_{\epsilon'}({\bf k}+{\bf q}')G^{R}_{\epsilon'} ({\bf k}+{\bf q}). 
\label{H1/2}
\end{eqnarray}
The calculation of the above diagrams in the diffusive limit
(\ref{diflim}) (the corresponding integrals 
are provided in Appendix A), gives 
\begin{eqnarray}    
h({\bf q},{\bf q}',\omega)&=&\frac{4\pi\tau^{4}}
{\hbar^{3}\Delta}[D(q^2+q'^{2})-i\omega].  \nonumber
\end{eqnarray}    

Having the basic ingredients of the disorder diagrammatics, we 
turn now to calculate the two-level correlation function 
defined by Eq.~(\ref{Rw}). Using the relation $\rho(\epsilon)=  
\mbox{Im} \left\{  \mbox{Tr} G^{R}(\epsilon) \right\}/\pi$, we have
\begin{eqnarray}    
R(\omega)=\frac{\Delta^{2}} {2\pi^{2}}  & \mbox{Re} & \left[ 
\left\langle \mbox{Tr} G^{R}(\epsilon+\omega)  \mbox{Tr} 
G^{A}(\epsilon)\right\rangle \right. \nonumber \\
&-& \left. \left\langle \mbox{Tr} G^{R}(\epsilon+\omega) 
\right\rangle \left\langle \mbox{Tr} 
G^{A}(\epsilon)\right\rangle \right]. \nonumber
\end{eqnarray}    
This formula can be used as a starting point 
for diagrammatic expansion. However, it 
produces a large number of diagrams. A convenient way of 
reducing this number is to express $R(\omega)$ in terms of 
a generating function which has a simpler diagrammatic expansion. 
This generating function, $F(\omega)$,  
has been found by Smith, Lerner and  
Altshuler\cite{Smith98}. It satisfies the relation: 
\begin{eqnarray}    
 R(\omega)&=&-\frac{\Delta^{2}}{2\pi^{2}\hbar^{2}}
\frac{\partial^{2}}{\partial \omega^{2}} 
\mbox{Re} F(\omega),\label{eq:RF}    
\end{eqnarray}    
and has the form of a free energy.
The diagrammatic expansion of $F(\omega)$ 
can be loosely pictured as an expansion in the number of 
diffusons and Cooperons loops: 
\begin{equation}    
F(\omega)=F_{1}(\omega)+F_{2}(\omega)+ F_3(\omega) + \cdots.
\label{eq:free}    
\end{equation}    
Thus, the leading term, $F_{1}(\omega)$,  
is the contribution of the one loop diagram (see   
Fig.~\ref{fig:free1}), $F_{2}(\omega)$ is the two-loop 
contribution (plus two additional terms whose role is 
to remove the ultraviolet divergence in the first diagram), 
$F_{3}(\omega)$ comes from  three-loop diagrams, etc\cite{Smith98}.  
In the periodic orbit picture, $F_1(\omega)$ is the contribution of
orbits shown in Fig.~1a+b, while $F_2(\omega)$ is, 
in essence, the contribution of the eight-shape orbits 
illustrated in Fig.~1c. 

The small parameter of the loop expansion (\ref{eq:free})
is $1/g$,  where $g$ is the dimensionless conductance of the system.
$g \propto t_H/t_c$ is the ratio of the Heisenberg time, 
$t_H= 2 \pi \hbar/\Delta$, to the classical relaxation time of particles 
in the system, $t_c$. In diffusive systems, $t_c=L^2/D$ 
(known as the Thouless  time\cite{Thouless74}) is the time
 which takes for a classical 
particle to diffuse  across the system.

{\narrowtext    
\begin{figure}      
  \begin{center}     
\leavevmode     
        \epsfxsize=7.33cm        
         \epsfbox{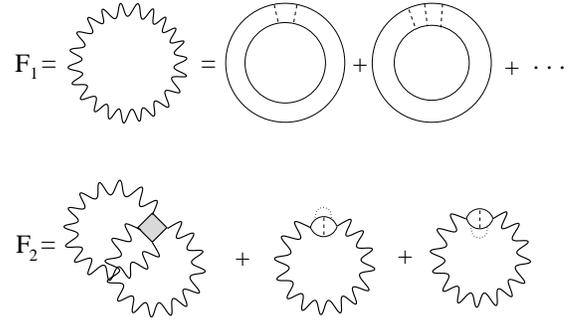}      
\end{center}     
   
	\caption{ Diagrams of
the free energy:  $F_1(\omega)$ is the leading 
contribution associated with
the diagonal approximation of periodic orbit theory. 
$F_{2}(\omega)$ is the WL contribution to the free energy associated 
with the eight-shaped orbits of Fig.~1c. Dashed impurity lines represent
large momentum transfer, $k> 1/l$, dotted lines represent small momentum
transfer, $k< 1/l$. }    
	\label{fig:free1}    
\end{figure}    
} 
The form of the free energy (\ref{eq:free}) together with
Eq.~(\ref{eq:RF}) induces a similar expansion for 
the two-level correlator: 
\begin{eqnarray}
R(\omega) = R_1(\omega) + R_2 (\omega) + R_3 (\omega) + \cdots, \nonumber
\end{eqnarray}
where
\begin{eqnarray}    
 R_j(\omega)&=&-\frac{\Delta^{2}}{2\pi^{2}\hbar^{2}}
\frac{\partial^{2}}{\partial \omega^{2}} 
\mbox{Re} F_j(\omega),  ~~~~j=1,2,3 \cdots.\label{eq:RFi}    
\end{eqnarray}    
Thus $R_1(\omega)$ is the result of diagonal approximation,
$R_2(\omega)$ is the WL contribution, 
and additional terms give  higher WL corrections.

The leading contribution to the two-level correlation function,
$R_1(\omega)$, has been discussed extensively  
by Altshuler and Shklovskii\cite{altshulershklovskii}.
It is straightforwardly calculated using (\ref{eq:RFi}).
Taking into account the $1/n$ symmetry factor of 
the $n$ ladder diagram defining $F_1(\omega)$ (see Fig.~9) 
we obtain  $F_{1}(\omega)=-\sum_{q}\ln(Dq^{2}\tau-i\omega\tau)$,
where the diffusive approximation (\ref{diflim}) has been assumed.  
Notice that although this sum does not converge, its second  
derivative with respect to $\omega$ does.
Moreover, one can check that, in two dimensions,
$R_1(\omega) =0$ for  $\omega> 0$. Thus the leading 
term in this case is the WL contribution\cite{Kravtsov95}. 
 
In this paper we focus our attention on the WL contribution to $R(\omega)$ 
of two dimensional ballistic systems. As a reference point, 
however, it will be instructive to review results of RMT, and
disorder diagrammatics in the diffusive limit. 
In both cases, our starting point is the diffusive form of the 
WL contribution to the free energy (obtained from the diagrams 
shown in Fig.~9): 
\begin{eqnarray}    
F_{2}(\omega)&=&\frac{\Delta}{\hbar\pi}\sum_{{\bf q},{\bf q}'} 
\frac{i\omega}{(Dq^{2}-i\omega)( Dq'^{2}-i\omega)}.\label{f2dif}    
\end{eqnarray} 

The RMT result corresponds to the zero mode contribution (${\bf q}={\bf q}'=0$)
in the above sum, namely 
$F_2(\omega) \simeq -i \Delta/ \hbar \pi \omega$.  It is
purely imaginary, therefore, Eq.~(\ref{eq:RFi}) implies  
that $R_2(\omega)=0$ in the RMT limit. Since RMT accounts for the universal
behavior of chaotic system, we conclude that $R_2(\omega)$
is a purely nonuniversal quantity. 

Turning to the diffusive limit, we first note that 
the sum in (\ref{f2dif}) diverges logarithmically, even 
after differentiating twice 
with respect to $\omega$. Thus one has to introduce an upper cutoff 
on the momentum, which is usually taken to be $1/l$, where $l$ is the elastic 
mean free path. As will be shown in the next chapter this 
artificial cutoff can be avoided if the approximations
(\ref{diflim}) are not used in the calculations.
    
To evaluate $F_{2}$ in the regime $1/t_c \ll \omega \ll 1/\tau$,
one can also use dimensional regularization\cite{Smith98}: 
Replacing the sums over ${\bf q}$ and ${\bf q}'$  
by integrals, and evaluating them in $d=2+\eta$ dimensions yields
\begin{eqnarray} 
F_{2}(\omega)&=&\frac{i\omega L^{4}\Delta}{\pi\hbar}\left[\int\frac{d^{d}q} 
{(2\pi)^{d}}\frac{1}{ Dq^{2}-i\omega}\right]^{2}. \label{dimreg}
\end{eqnarray} 
Changing variables from $q$ to $(iD/\omega)^{1/2}q$ and using the 
formula (see Appendix A) 
\begin{eqnarray} 
\int\frac{d^{d}q}{1+q^{2}}=\pi^{d/2}\Gamma(1-\frac{d}{2}), \nonumber
\end{eqnarray} 
one arrives at 
 \begin{eqnarray} 
F_{2}(\omega)
&=&\frac {iL^{4}\Delta}{\hbar\pi(4\pi D)^{2}}
\left(\frac{-i}{4\pi D}\right)^{\eta}\Gamma^{2}(\frac{-\eta}{2})
\omega^{1+\eta} .\nonumber
\end{eqnarray}  
$R_2(\omega)$ is now obtained by taking 
the second derivative with respect to $\omega$, 
as follows from Eq.~(\ref{eq:RFi}). Thus
using $\Gamma(1+\eta/2)\Gamma(-\eta/2)=\pi/\sin(-\eta\pi/2)$ we have
\begin{eqnarray} 
R_{2}(\omega)
&=&- 
\frac{\Delta\pi(1+\eta)\eta \mbox{Re}[(-i\omega)^{\eta-1}]}{2\hbar g^{2}
(4\pi D)^{\eta} \sin^{2}(-\eta\pi/2)
\Gamma^{2}(1+\eta/2)}, \nonumber
\end{eqnarray} 
where $g=4 \pi^2 \hbar D/L^2 \Delta$ is 
the dimensionless conductance of the system.
Finally we let $\eta \to 0$, and obtain
\begin{eqnarray}    
R(\omega)\simeq R_2(\omega)=-\frac{\Delta} { g^{2}\hbar \omega},\hspace{1cm}
\frac{1}{t_{c}}\ll \omega \ll \frac{1}{\tau}. \nonumber
\end{eqnarray}   
Note that the domain of validity of the above formula
vanishes in the ballistic limit since the classical relaxation time, $t_c$, 
is smaller or equal to the scattering time,  $\tau$.

\section{Weak localization in the nondiffusive regime}   
In this section we calculate the WL contribution to
the two-point correlator in the ballistic regime. 
By ballistic we refer to the situation in which the elastic 
mean free path, $l$, is of order of the size of the system $L$.
To understand what kind of changes are needed in order 
to extend the diagrammatic calculation into the ballistic regime,
recall that the diffusive approximation (\ref{diflim}) corresponds to
the leading order result in the small parameter $l/L$ (since
$ql \ll 1$, $q$ is of order $1/L$, and $l \ll L$). In the ballistic 
regime this approximation cannot be used, and one has to  
evaluate integrals, such as (\ref{eq:Pi}) and (\ref{Hbox}), 
to all orders in $l/L$. Moreover,
diagrams having a small number of impurity lines form the 
dominant contribution (unlike in the diffusive regime), 
therefore, possible cancellations
among diagrams as well as double counting should 
be examined carefully. The outcome of this examination
is that  diffusons and Cooperons contributing to $F_2(\omega)$
should start from two impurity lines. 
Apart form this point, the  WL contribution 
is given by the same diagrams shown in Fig.~9, but evaluated
to all orders in $l/L$.

We begin by deriving an expression for the diffuson (starting from
two impurity lines) in the ballistic regime. Dyson's equation, 
in this case, yields  
\begin{eqnarray}   
{\cal D}({\bf q},\omega)&=&\frac{\hbar}{2\pi\nu\tau}
\frac{\Pi(\omega,{\bf q})}{1-\Pi(\omega,{\bf q})},  \nonumber
\end{eqnarray}    
where $\Pi(\omega,{\bf q})$ is the integral given by Eq.~(\ref{eq:Pi}).
To avoid the expansion in $ql$ and $\omega\tau$, here we first
integrate over $\xi$ (by closing the contour in the complex plain), 
and then integrate over the angle, $\theta$, exactly. The result is 
$\Pi(\omega,{\bf q})=1/Q_{\omega}(q)$, where 
\begin{eqnarray}   
 Q_{\omega}(q)&=&\sqrt{(1-i \omega \tau)^{2}+(lq)^{2}}.  \nonumber 
\end{eqnarray}   
Thus the generalized formula for the diffuson is
\begin{eqnarray}   
{\cal D}({\bf q},\omega)  
&=&\frac{\hbar}{2\pi\nu\tau}\frac{1}{Q_{\omega}(q)-1}.\label{eq:exactdiffuson} 
\end{eqnarray}   
A similar calculation for the 
Cooperon produces the same analytical expression.    

The above formula holds both in the ballistic as well as the diffusive 
regime (\ref{diflim}).
It can be easily checked that an expansion of the denominator of
Eq.~(\ref{eq:exactdiffuson})
in $\omega \tau$ and $ql$, yields the result
of the diffusive limit, (\ref{diffuson}). 

The calculation of the Hikami box (Fig.~8b, Eq.~\ref{Hbox}), 
in the ballistic limit, follows along the same lines. Namely, 
one first integrates over the modulus of ${\bf k}$, 
and then the remaining angular 
integration is performed exactly. For example, after integration
over the modulus of ${\bf k}$,  Eq.~(\ref{H1/2}) reduces 
to an integral of the from
\begin{eqnarray}   
{\cal I}(x_1,x_2,\varphi)=\frac{1}{2\pi}\!\! \int_{0}^{2\pi} \!\!\!\! 
\frac{d \theta}{(1\!+\!x_1\cos\theta)(1\!+\!x_2\cos(\theta\!-\!\varphi))}.   
 \label{eq:int1}  
\end{eqnarray}   
The result of the integration over the angle $\theta$ (see Appendix A)
is
\begin{eqnarray}  
{\cal I}(x_1,x_2,\varphi)= \left(\frac{1}{y_1}+\frac{1}{y_2} \right)
 \frac{1}{1+y_1     
   y_2-x_{1}x_{2}\cos(\varphi)}    \label{Iresult}
\end{eqnarray}   
where  

{\narrowtext    
\begin{figure}      
  \begin{center}     
\leavevmode     
        \epsfxsize=7.33cm        
         \epsfbox{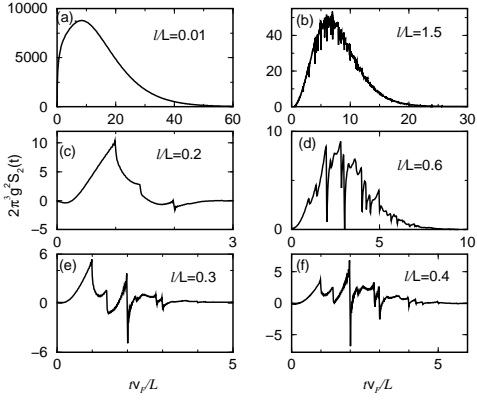}      
\end{center}     
   
	\caption{The results for the WL contribution to the 
structure  factor at various values of the ratio of the elastic mean free 
path, $l$, to the size of the system $L$. 
The WL effect, in this system, becomes stronger in two limits:   
(a) the diffusive regime, $\l \ll L$, where the particle approaches
localization in real space, and (b) the  ballistic limit, $l\gg L$, 
where the particle becomes localized in  momentum space.}   
\label{fig:results}   
\end{figure}	   
} 
\noindent
\begin{eqnarray}  
y_i= \sqrt{1-x_{i}^{2}},~~~~~~i=1,2. \nonumber  
\end{eqnarray}  

With the help of this function, the various terms of the Hikami box 
(see Eq.~(\ref{Hbox}) and Fig.~8b) take the form:  
\begin{eqnarray}  
h_{\pm}  = -\frac{2\pi \tau^{3}} {\hbar^{3}(1-i\omega\tau)^{4} \Delta}   
 {\cal I}^2 \left[\frac{ilq_{1}}{1-i\omega\tau},     
\frac{\pm ilq_{2}}{1-i\omega\tau},\varphi_{12}\right],  \nonumber   
\end{eqnarray}  
where $\varphi_{12}$ is the angle between    
${\bf q}_1$ and ${\bf q}_2$,  
and  
\begin{eqnarray}  
h_{0} =\frac{4\pi\tau^{3}(1\!-\!i\omega\tau)}   
{\hbar^{3}f_1^2 f_2^2 \Delta } {\cal I} \left[\left( \frac{lq_{1}}    
{\sqrt{2}f_1 }\right)^{2}\!,     
\left(\frac{lq_{2}} {\sqrt{2}f_2}\right)^{2}\!, \nonumber  
2\varphi_{12}\right],    
\end{eqnarray}  
where  
\begin{eqnarray}  
f_i= \sqrt{(1-i\omega \tau)^2 + (q_i l)^2/2}~~~~~~i=1,2.\nonumber  
\end{eqnarray}  
Collecting the diagrams of $F_2(\omega)$ (Fig.~9) we obtain  
\begin{eqnarray}  
F_{2}(\omega)=\frac{\hbar^2 \!\Delta^2}{4 \pi^2 \! \tau^2}\! \sum_{1,2}   
\!{\cal D}_1  {\cal D}_2 \! \left[  
h_0\!+\! (h_{+}\!+\!h_{-})\! \left(\! \frac{1}{{\cal D}_1}\!+\!   
\frac{1}{{\cal D}_2}\!+\! 1\!\right) \right], \nonumber
\end{eqnarray}  
where we use the notation 
\begin{eqnarray}
{\cal D}_i=2 \pi \nu \tau   
{\cal D}({\bf q}_i,\omega)/\hbar, \nonumber
\end{eqnarray}  
and the sum is over the vectors ${\bf q}_1$ and ${\bf q}_2$.  
The periodic boundary conditions in our system imply that  
${\bf q}_i\!=\! 2 \pi {\bf m}_i/L$ where ${\bf m}_i$ is an 
integer vector of two components.   

The above formula is our central result. 
Performing the sum over momenta 
and substitution it in (\ref{eq:RFi}) gives the exact WL contribution 
to the two-level correlation function in the semiclassical limit.  The 
applicability range 
{\narrowtext    
\begin{figure}    
\vspace{-2.5 cm}  
  \begin{center}     
\leavevmode     
        \epsfxsize=7.33cm        
         \epsfbox{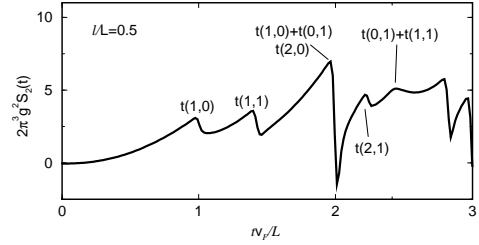}      
\end{center}     
   
	\caption{The weak localization contribution to the structure factor, 
in the ballistic regime, exhibits a singular behavior. 
The singularities are located
at times which are linear combinations of periods of two orbits 
 of the clean system (see Fig.~4). 
Here $t(n_x,n_y)=(n_x^{2}+n_y^{2})^{1/2}L/v_{F}$ denote the period of the
periodic orbit defined by the pair of winding numbers $(n_x,n_y)$.}
\label{fig:resultswnum}  
\end{figure}	   
}

\noindent
of our result goes beyond the 
diffusive limit, $l\ll L$, and includes
the ballistic regime, $l \sim L$, as well.
In contrast with the formula in the diffusive limit (Eq.~\ref{f2dif}), 
here the momenta sum converges, and there is no need to introduce 
an arbitrary cutoff or regularization. The results which will be shown
below were obtained by performing the momenta sum 
numerically with cutoff chosen such that  contribution 
of additional terms is of order of the numerical error. 

In presenting our results it will be convenient
to employ the  spectral structure factor defined in Eq.~(\ref{eq:sdef}). 
We denote by $S_2(t)$ the corresponding WL contribution, 
\begin{equation} 
S_2 (t) = \frac{\hbar}{\Delta}\int d \omega R_2(\omega) e^{- i \omega t},  
\end{equation} 
and rescale its magnitude by a factor of $2 \pi^3 g^2$, where 
$g \propto t_H/t_c$ is the dimensionless conductance of the system. 
Note that in the ballistic regime,
the relaxation time, $t_c$, is no longer the
diffusion time. 
It  is approximately the traversal time across the system,  $t_c=L/v_{F}$ 
where $v_{F}$ is the velocity of the particle, and $L$ is the size 
of the system. Therefore, from now on we define $g$ to be
\begin{equation}
g=\frac{\hbar v_{F}}{\Delta L}.
\end{equation}

In Fig.~\ref{fig:results} we plot $S_2 (t)$, for
various values of the ratio between the elastic mean
free path and the size of the 
system. These values range 
from diffusive ($l/L=0.01$) to ballistic ($l/L=1.5$) dynamics. 
Several features of $S_2(t)$ are evident: First, the WL
contribution appears only within a finite interval of time. It vanishes 
both at $t=0$ and when $t \to \infty$. Second, in both limits,  
$l\ll L$ and $l \gg L$, the WL contribution becomes strong. 
Third, 
{\narrowtext    
\begin{figure}     
\vspace{-5.5 cm}  
  \begin{center}     
\leavevmode     
        \epsfxsize=7.33cm        
         \epsfbox{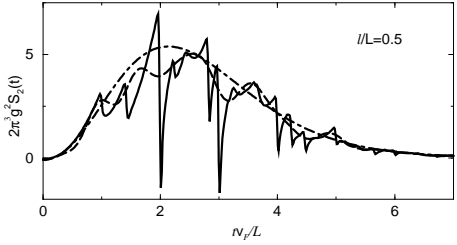}      
\end{center}     
   
	\caption{The convergence behavior of the momenta sum of the 
structure factor in the ballistic regime. 
The smooth part (dash-dotted) is determined by the zero mode, 
${\bf q}_1={\bf q}_2=0$. The dashed line is the result 
of a sum over momenta within radius of $2.83 \pi$.
Higher $q$ terms  build up singularities 
along the time axis as demonstrated by the solid line where 
the momenta sum extends to radius of $80 \pi$. }   
	\label{fig:exactdigrams1}   
\end{figure}
}   
\noindent 
in the ballistic regime, $l \sim L$, $S_2(t)$ 
exhibits a distinctive singular behavior 
consisting of a series of dips. These dips are located
at times which are  
combinations of periods of the periodic orbits of the clean system.  
In Fig.~\ref{fig:resultswnum} we depict $S_2(t)$ for 
$l=L/2$ and indicate the singularities with the 
corresponding combinations of periodic orbits.

\section{Analysis}
 
In analyzing the above results it is instructive to study, first, 
the convergence behavior of the momenta sum 
of the WL contribution. In Fig.~\ref{fig:exactdigrams1} we depict the results 
for $S_2(t)$ calculated in the following approximations: The 
dash-dotted line 
is the contribution coming from the zero mode, ${\bf q}_1={\bf q}_2=0$. 
Clearly this mode dictates the gross behavior of $S_2(t)$. 
In particular, it determines the interval of time where WL is 
significant. The dashed line 
is the result obtained by taking 
into account the next lowest
 momentum modes, i.e. summing over ${\bf q}_1$ and ${\bf q}_2$ 
within the radius $q_1,q_2 \leq \sqrt{8} \pi$. In this approximation  
some additional features of $S_2(t)$ are resolved. The solid line is
the result of the full momenta sum. Thus the singular
behavior of the structure factor comes from the tail of the sum. 

To obtain a simple analytic characterization of the WL 
contribution to the structure factor, we 
proceed in the following way. First, we derive a formula 
for the smooth part of $S_2(t)$ given by the contribution of the zeroth mode 
${\bf q}_1={\bf q}_2=0$. This formula will give us the main parameters 
characterizing the WL contribution in the ballistic limit.
Then, we evaluate the momenta sum in the asymptotic limit of
large $\omega$. The result of this calculation 
provide the local behavior of $S_2(t)$ in the vicinity 
of the singularities.  
 
To calculate the smooth part of $S_2(t)$, denoted hereinafter 
by $\bar{S}_{2}(t)$, we start by evaluating the zero mode
contribution to $R_2(\omega)$. A straightforward calculation 
of the term ${\bf q}_1={\bf q}_2=0$ yields
\begin{eqnarray} 
R_2(\omega; q=0~\mbox{contrib.}) = \frac{ 4 \tau^3 \Delta^{3}  
\left[ 5 -54 (\omega \tau)^2 +21(\omega\tau)^4\right]}{\hbar^{3}\pi^3 [1+  
(\omega \tau)^2]^6}.  \nonumber
\end{eqnarray} 
Taking, now, the Fourier transform  we obtain: 
\begin{equation}  
\bar{S}_2(t) = \frac{l^{2}}{12\pi^{2} g^{2}L^{2}} e^{-\tilde{t}} ~\tilde{t}^{2}
(\tilde{t}^{3}+3\tilde{t}^2 + 6\tilde{t} + 6),
\label{envelope} 
\end{equation} 
where $\tilde{t}= t/\tau$. We remark, here, that the above
formula applies only in the ballistic 
regime, where the zeroth mode is dominant. 
In the diffusive systems, the zero mode contribution 
is negligible compared to that coming from the momenta sum in
regime $1/L < q \ll 1/l$ (which is absent in the ballistic case).

Formula (\ref{envelope}) allows one to characterize the major 
features of the WL contribution to the structure factor:  
The time $t^*$ where  $\bar{S}_2(t)$ is maximal; its 
value at this point, $S_2^*$; and the width of the time interval
where the WL effects are appreciable, $W^*$. The results are: 
\begin{eqnarray} 
t^* &= &4.24~ \tau, \nonumber \\ 
W^* &=& 2.46 ~\tau, \nonumber 
\end{eqnarray}  
and 
\begin{eqnarray} 
S_2^* &= &0.353~\left(\frac{l}{Lg}\right)^2=
0.353~\left(\frac{ \tau\Delta}{\hbar}\right)^2,
\nonumber
\end{eqnarray}  
where the interval width is defined by 
$(W^*)^2= \int d t \bar{S}_2(t) (t-t^*)^2/
\int d t \bar{S}_2(t)$. Thus WL effects, in the 
ballistic limit, are pronounced within a time interval
of width $2.46 \tau$
centered at $t=4.24 \tau$, and the typical value of the WL 
contribution is proportional to $\tau^2$. 

Note that these results are independent of the size of the system.
Therefore the gross behavior of $S_2(t)$ is not influenced by the
periodic orbits of the clean system.
It is natural to ask what is the role 
the classical orbits of the system? 
As we show below, these orbits lead to the singular features
decorating the smooth part of the structure factor $\bar{S}_2(t)$ 
as demonstrated in Fig.~12. 

In analyzing this singular behavior, we first notice that its
main part comes from large $\omega$ or equivalently
large values of the momenta ${\bf q}_1$ and ${\bf q}_2$. 
Therefore, to calculate this contribution it is sufficient to 
approximate the discrete angular sum of $F_2(\omega)$
(over the phase between the vectors  ${\bf q}_1$ and ${\bf q}_2$)
by an integral. The small parameter of this approximation is 
$1/\omega \tau$. 
From (\ref{Iresult}) one finds that the angular average of
the WL contribution to the free energy, denoted by 
$\bar{F}_{2}(\omega)$,
is:
\begin{eqnarray}   
\bar{F}_{2}(\omega) =\frac{\tau\Delta} 
{\pi\hbar}\sum_{{\bf q}_{1},{\bf q}_{2}} 
\frac{A+B+C} {Q_1Q_2   
 (Q_1+Q_2)(Q_1-1)(Q_2-1)},  \nonumber 
\end{eqnarray}   
where   
\begin{eqnarray}   
A=\frac{1}{1-i\omega\tau},~~~B=\frac{1-i\omega\tau}   
{Q_1Q_2},~~~C=-B (Q_1+Q_2), \nonumber   
\end{eqnarray} 
and
\begin{eqnarray}
Q_i= Q_\omega(q_i). \nonumber   
\end{eqnarray} 
At asymptotically large values of  $\omega\tau$ the leading contribution
comes from $C$. This is evident once noticing that 
when $\omega \tau \to \infty$, $Q_i \to \omega \tau$, 
and therefore $A, B = O(\frac{1}{\omega \tau})$ whereas $C = O(1)$. 

Next we apply the Poisson summation formula to convert the sum over
${\bf q}_1$ and ${\bf q}_2$ into an integral. The free energy
is then  expressed as
 \begin{eqnarray}   
F_{2}(\omega)\simeq \bar{F}_{2}(\omega)  
 =\sum_{{\bf m},{\bf n}} F_2^{({\bf m},{\bf n})}, \label{sumterms}
\end{eqnarray}   
where ${\bf m}$ and ${\bf n}$ are integer vectors.
As will be shown below, these integer vectors are associated with  
winding numbers of the periodic orbits of the clean system. 
Each term in formula (\ref{sumterms}) is of the form
\begin{eqnarray}   
F_2^{({\bf m},{\bf n})}=-\frac{\tau \Delta L^{4}}{\pi\hbar} 
(1-i \omega \tau) K(m)K(n),   
\label{f2mid}
\end{eqnarray}   
where   
\begin{eqnarray}   
K(m)=\int_{0}^{\infty}\frac{dq}{2 \pi} \frac{qJ_{0}(mqL)
}{Q_{\omega}^{2}(q)(Q_{\omega}(q)-1)}.   \label{eq:km}
\end{eqnarray}  
Here $J_0(x)$ is the Bessel function of zero order, and $m=|{\bf m}|$ 
is the magnitude of the vector ${\bf m}$. For $m=0$ this integral yields
\begin{eqnarray}   
K(0)= \frac{1}{2 \pi l^2(1- i \omega \tau)} +
 O\left(1/\omega^2\right). \nonumber
\end{eqnarray}
For $m \neq 0$ the integral (\ref{eq:km})
can be  calculated using the steepest descents method 
(see Appendix B), and in the large $m$ limit it gives
\begin{equation}
K(m)\sim -\frac{e^{3/2}}{\sqrt{72}\pi l^{2}}
\frac{e^{-\frac{mL}{l}(1-i\omega\tau)}}{1-i \omega\tau}.\label{eq:Kresult}
\end{equation}

The above results imply the following form of the structure factor
\begin{eqnarray}
S_2(t)\sim \bar{S}_2(t) + \sum_{{\bf n},{\bf m}}  
S_2^{({\bf n},{\bf m})}(t), \nonumber
\end{eqnarray} 
where
\begin{eqnarray}  
S_2^{({\bf n},{\bf m})}(t) = 
- B_{nm}\left(\frac{L}{gl} \right)^2 
\Theta[t-t_{nm}]~\tilde{t}^2 e^{-\tilde{t}}   \nonumber  
\end{eqnarray}  
is the contribution associated with orbits   
characterized by the winding vectors ${\bf n}$ and ${\bf m}$.   
Here $\tilde{t}=t/\tau$, $t_{nm}= (n+m) L/v_F$, and $\Theta(x)$ is
the step function. The amplitude of each contribution, $B_{nm}$, depends
on the values ${\bf n}$ and ${\bf m}$. For cases where
either ${\bf n}$ or ${\bf m}$ vanish, $B_{0m}=B_{n,0} \simeq e^{3/2}/12 
\sqrt{2} \pi^4$, while  if $n$ and $m$ are large, $B_{nm} \simeq
e^3/ 144 \hbar^2 \pi^{4}$.

Thus, $S_2(t)$ is composed of a smooth contribution  (\ref{envelope}) 
and a sequence of singular functions, 
of the form $-t^2 e^{-(t-t_{mn})/\tau}\Theta(t-t_{mn})$, 
where $t_{mn}$ is the period     
of a composite orbit, i.e. the sum of periods of two periodic orbits of the 
clean system. Each singular contribution is negative, and its
magnitude at time $t_{mn}$ is proportional to 
$ t_{mn}^2 e^{-t_{mn}/\tau}/l^{4}$. The contribution associated with
single orbits (i.e.~when either $n$ or $m$ vanish) is considerably larger
than that of composite orbit (in which both $n$ and $m$ differ from zero).
In any case, the
singular contribution decreases exponentially in time , 
and as a power law in $1/l$ (when  $L<l$). This behavior is indeed
observed in Figs.~10, 11, and 12.

\section{Summary and concluding remarks}

In this paper we have calculated the WL contributions to 
the two-level correlation function and its Fourier 
transform, the structure factor.
These are the leading quantum interference effects 
which affect the spectral statistical properties of 
the system defined in (\ref{eq:hamiltonian}). 

Our theory generalize previous calculations 
of the  WL contribution to the spectral statistics of diffusive
systems\cite{Kravtsov95} by extending them into the ballistic
regime where the elastic mean free path, $l$, is of order of 
the size of the system, $L$.
Here the disorder is weak enough to leave 
traces of the dynamics of the underlying clean system, which
appear as singularities 
in the structure factor (Figs.~\ref{fig:results}, \ref{fig:resultswnum}, 
and \ref{fig:exactdigrams1}).

Our study has been focused on spectral rather than 
dynamical characteristics to avoid the problem of specifying
the manifold on which WL takes place. Indeed
Fig.~\ref{fig:results} demonstrates that the WL contribution
is pronounced in two limits.
Panel (a) of Fig.~10 is a representative 
example of the results deep in the 
diffusive regime $l\ll L$, while 
panel (b) shows the typical behavior in the ballistic limit, $l=1.5 L$. 
In both cases, the system approaches the strong localization limit, but 
the localization is of different nature. In the diffusive case, it is 
localization in real space \cite{Anderson58}, whereas in the ballistic 
case the 
localization is  on a quasi-one-dimensional annulus in the momentum space.
(This is evident once noticing that on clean torus, 
eigenstates are plain waves and therefore the particle is localized in 
momentum space.) In the latter case, it is suggestive 
that the effective dynamics  is associated 
with Levy flights \cite{Mandelbrot82} rather than diffusion, since the disorder
couples, predominantly, momentum states with 
degenerate eigenvalues, which may lie far away along the 
momentum annulus\cite{Altshuler97}.

A simple semiclassical interpretation of our results, within periodic 
orbit theory, is not straightforward. The results, clearly, cannot be obtained
from a diagonal approximation in which higher order $\hbar$ corrections 
are added to Gutzwiller's trace formula (e.g.~diffracting orbits, 
creeping orbits, etc.~). One can easily verified that such approximation 
yields only a positive contribution, in contrast with our results.
This type of correction might explain the smooth positive part of the
WL contribution, $\bar{S}_2(t)$.
However, a correct analysis within periodic orbit theory
must go beyond the diagonal approximation, and take into account 
pairing of orbits which are not related by symmetry, but  have 
actions exponentially close one to the other (up to a constant 
phase, $\pi$, which is needed in order to explain the negative 
contribution of the periodic orbits). The fact that the WL contribution
may become negative at certain times implies that the system  
exhibits anti-weak-localization at certain regions in phase space.
This may be related to anti-scarring effects observed in wave functions
of chaotic billiards\cite{Agam94}.

Nevertheless, our work still elucidates several features of the leading
WL effects in ballistic chaotic systems.  First it shows that 
it appears within a finite interval of time;  Second, it has a 
singular behavior associated
with periodic orbits and linear combinations of periodic orbits; Third
it can have different signs at different points in phase space.

These results have important consequences: First they show that
the dominant contribution to the WL, in the ballistic regime, 
does not come from the eight-shaped orbits (Fig.~1c), as suggested by
the diagrammatic picture. The main contribution comes from
diffracting orbits (which are not related to the classical periodic
orbits of the system), as well as from 
the original periodic orbits of the system. 
Moreover, the zero mode contribution, defining $\bar{S}_2(t)$, 
plays a dominant role here, while according to the results of
the ballistic $\sigma$-model it should vanish (since the zero 
mode of the $\sigma$-model is identical to RMT). The apparent contradiction
between our results and those of the ballistic $\sigma$-model 
is probably due to the fact that the ballistic $\sigma$-model
does not account correctly for the return probability. This is
also manifested by the so called ``repetition problem'' 
which is a small mismatch, associated with repetitions of 
periodic orbits, between the exact asymptotic results of 
periodic orbit theory and those of the ballistic $\sigma$-model.
Ideas associated with memory effects in long range random 
potential\cite{Wilke00} may be useful in resolving this 
problem.
\newline
\newline
\begin{center}
{\large \bf Acknowledgment}
\end{center}
\vspace{0.5cm}

It is our pleasure to thank Igor Aleiner, Alex Altland, Boris Altshuler,
John Chalker, Igor Lerner,  and Rob Smith for useful  
discussions. This work has been initiated at   
the ``Extended Workshop on Disorder, Chaos and Interactions in  
Mesoscopic Systems'' (Trieste, Aug. 98). We thank   
the I.C.T.P. and especially Volodya Kravtsov for the generous hospitality.
This research was supported by THE ISRAEL SCIENCE FOUNDATION founded by
The Israel Academy of Science and Humanities, and by
The Herman Shapira foundation.
\newline
\newline
\begin{center}
{\large \bf Appendix A. Useful integrals}\label{app:integral}
\end{center}
\vspace{0.5cm}

In this appendix we calculate useful integrals
frequently encountered when calculating diagrams which 
appear in this paper. The first type of such integrals appear when 
integrating products of retarded and advanced Green functions 
over the energy. The integral is of the form:
\begin{eqnarray}
Y_{n,m}=\int_{-\infty}^{\infty}d\eta 
\left(\frac{1}{\eta+\frac{i}{2\tau}}\right)^{n}
\left(\frac{1}{\eta-\frac{i}{2\tau}}\right)^{m},\nonumber                    
\end{eqnarray}
where $n$ and $m$ are non-zero integers. Applying the Cauchy theorem, 
and using the fact that the coefficient $a_{-1}$  
of a Laurent series, $\sum_l a_{-l}(z-z_0)^{-l}$ 
of a function with an $n$-th order pole is 
$a_{-1}=\frac{1}{(n-1)!}\frac{d^{n-1}}{dz^{n-1}}
\left[(z-z_{0})^{n}f(z)\right]_{z=z_{0}}$, one immediately gets  
\begin{eqnarray}
Y_{n,m}=\frac{2\pi (m+n-2)!~i^{m-n}{\tau}^{m+n-1}}{(n-1)!(m-1)!}. \nonumber
\end{eqnarray}

The second type of integral appears when integrating 
over momenta in $d$ dimensions, e.g. in the
calculation of the WL contribution to the free energy 
(\ref{dimreg}) in the diffusive limit. This family of integrals is of the form 
\begin{eqnarray}
I_{n,d}=\int \frac{d^{d}q}{(1+q^{2})^{n}}, \nonumber
\end{eqnarray}
where $n$ is an integer, $d$ is a real number, and  
$d^d q= d\Omega q^{d-1}$ denotes the measure in $d$ dimensions.
Since the integrand is independent of the angles, the angular integral
yields $\int d\Omega = d \pi^{d/2}/\Gamma(1+d/2)$. This formula 
should be understood as an analytic continuation of a function defined  on 
an infinite set of integer values of $d$. Changing the integration variable 
to $x=q^{2}$ yields
\begin{eqnarray}
I_{n,d}=\frac{\pi^{d/2}d } {2\Gamma(1+d/2)}\int_{0}^{\infty} dx 
\frac{x^{d/2-1}}{(1+x)^{n}} =~~~~~~~~~~~~~~~~\nonumber\\
\frac{\pi^{\frac{d}{2}}d }{2\Gamma(1+\frac{d}{2})}
\frac{(-1)^{n-1}\partial^{n-1}}
{(n\!-\!1)!\partial\beta^{n-1}} 
\left. \int_{0}^{\infty}\! \! \!d\xi
\int_{0}^{\infty}\! \!\! dx x^{\frac{d}{2}-1}
e^{-\xi (x+\beta)}\right|_{\beta=1}. \nonumber
\end{eqnarray}
Integrating over x, then changing the integration variable to 
$y=\xi \beta$, and integrating over $y$ we obtain
\begin{eqnarray}
I_{n,d}=\frac{\pi^{d/2}\Gamma(d/2)\Gamma(1-d/2)
(-1)^{n-1}}{\Gamma(n)\Gamma(1+d/2-n)}. \nonumber
\end{eqnarray}
As the last step, we use the relation $\Gamma(x+1)= x\Gamma(x)$ 
to simplify the expression. The result is:
\begin{eqnarray}
I_{n,d}=\pi^{d/2} \frac{\Gamma(n-d/2)}{\Gamma(n)}. \nonumber
\end{eqnarray}

We turn now to evaluate the integral which appears when 
calculating diagrams in the ballistic limit, namely
the integral defined by Eq.~(\ref{eq:int1}).  
It will be calculated in the regime $|x_{1}|,|x_{2}|<1$
(corresponding to the diffusive limit) and then analytically 
continued to the full complex plain.
It is natural to substitute $z =e^{i \theta}$ which 
immediately transforms (\ref{eq:int1})
into the contour integral
\begin{eqnarray}
I(x_1,x_2,\varphi) = ~~~~~~~~~~~~~~~~~~~~~~~
~~~~~~~~~~~~~~~~~~~~~~~ \nonumber \\
~~~~~=\frac{-2i}{\pi x_1x_2}\! \oint_{|z|=1}\! 
\frac{zdz}{(z^{2}\!+\!\frac{2}{x_1}z\!+\!1)
(z^{2}e^{-i\varphi}\!+\!\frac{2}{x_2}z\!+\!e^{i\varphi})} \nonumber \\
~~~~~=\frac{-2i e^{i \varphi}}{\pi x_1x_2} \oint_{|z|=1} \frac{zdz}
{(z-z_{1})(z-z_{2})(z-z_{3})(z-z_{4})}\nonumber
\end{eqnarray}
where $z_{i}$ are
\begin{eqnarray}
z_{1}=-\frac{1}{x_1}+\sqrt{\frac{1}{x_1^{2}}-1}, ~~~~~~~~~
 z_{2}=\frac{1}{z_{1}},\nonumber\\
z_{3}=-\left(\frac{1}{x_2}+\sqrt{\frac{1}{x_2^{2}}-1}\right)e^{i \varphi}, 
~~~~~~~ z_{4}=\frac{e^{i 2 \varphi}}{z_{3}}.\label{zis}
\end{eqnarray}
Only the poles $z_{1}$ and $z_{3}$, which lie inside the unit circle,
contribute to the integral. Thus using
the residue theorem we get

\begin{eqnarray}
I(x_1,x_2,\varphi)=\frac{4}{x_1x_2}\left[\frac{z_{1}}{(z_{1}-z_{2})
(z_{1}-z_{3})(z_{1}-z_{4})} \nonumber \right. \\
\left. +
\frac{z_{3}}{(z_{3}-z_{1})(z_{3}-z_{2})(z_{3}-z_{4})}\right]~~~~~~~~~~~~~
\label{residue}
\end{eqnarray}
Finally, we substitute (\ref{zis}) in (\ref{residue})
and arrived at (\ref{Iresult}).
\newline
\newline
\newline
\begin{center}
{\large \bf Appendix B. Asymptotics of $K(m)$} 
\end{center}
\vspace{0.5cm}

In this appendix we evaluate the integral (\ref{eq:km}) 
in the asymptotic limit $m \gg 1$ and $\omega \tau \gg 1$. 
We begin by changing variables to $\eta=\frac{lq}{i+\omega\tau}$  
and taking the leading term in $1/(i+\omega\tau)$. The result is
\begin{eqnarray} 
K(m)&\simeq&\frac{1}{2 \pi l^{2}(i+\omega\tau)} 
\int\frac{\eta J_{0}(\frac{mL}{l}(i+\omega\tau) 
\eta)d\eta}{(\eta^{2}-1)^{\frac{3}{2}}}. \nonumber 
\end{eqnarray} 
Being interested in the leading order expansion in $1/\omega\tau$,
we further approximate the integral by substituting the 
asymptotic formula of the Bessel function: $J_{0}(x)
\simeq\sqrt{\frac{2}{\pi x}}\cos(x-\pi/4)$ as $x \to \infty$.
Representing the cosine as a sum of two exponents we arrive at:

{\narrowtext    
\begin{figure}      
  \begin{center}     
\leavevmode     
        \epsfxsize=7.33cm        
         \epsfbox{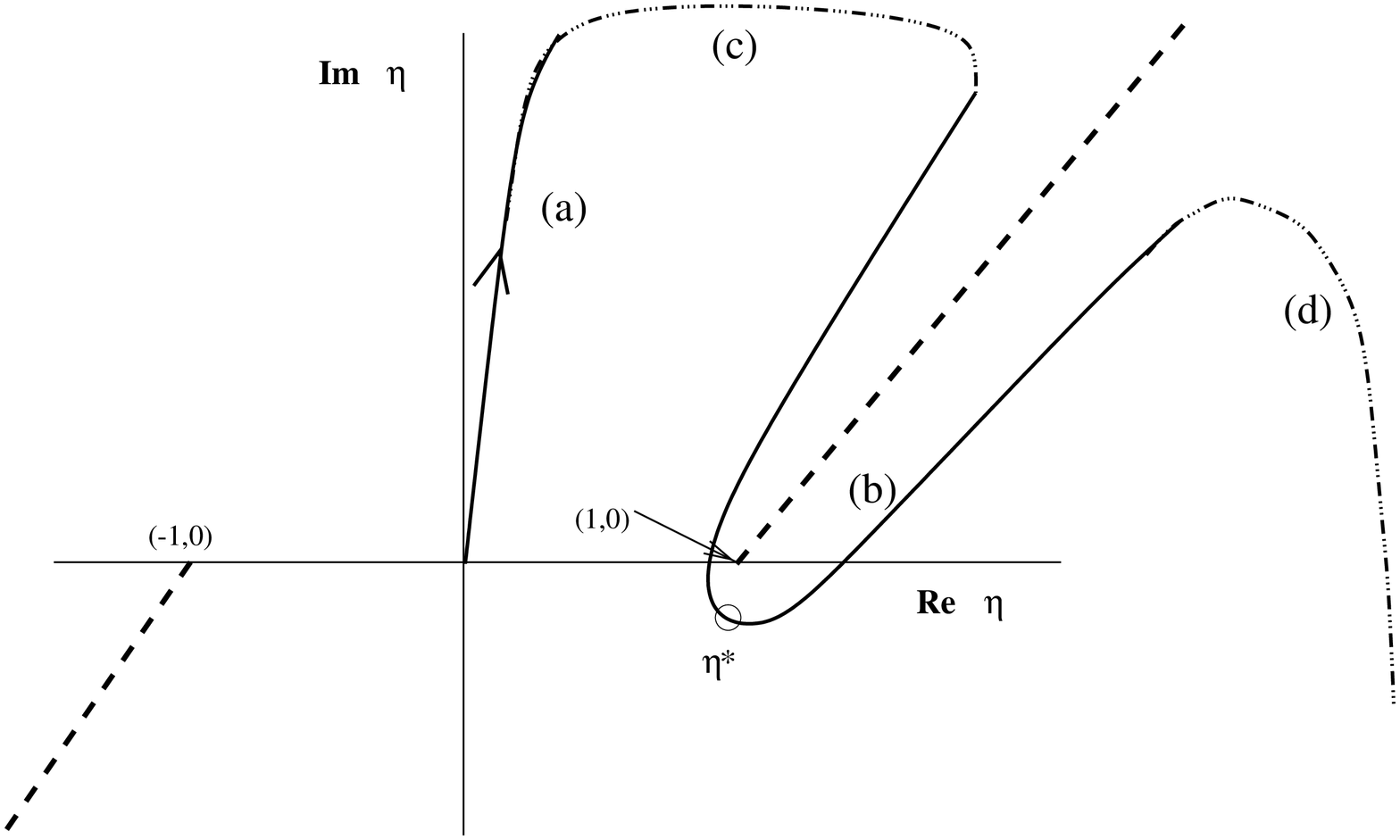}      
\end{center}     
   
	\caption{The contour of integration for the asymptotic evaluation of
$K(m)$. It is composed of steepest descent paths (a) and (b), connected by
arcs, (c) and (d), which give vanishing contributions. The dashed 
lines represent cuts of the integrand in the complex plain 
(see Eq.~\ref{eq:beforethetwist})}. 
\label{fig:cutss}  
\end{figure}  
}
\noindent
\begin{eqnarray} 
K(m) \simeq \frac{1}{
\sqrt{m L (2\pi l)^{3} (i+\omega\tau)^3 }}~~~~~~~~~~~~  \nonumber\\
~~~~ \times \sum_\pm 
\int \frac{\eta d \eta  }{(\eta^{2}-1)^{\frac{3}{2}}}
e^{\pm i\left[\frac{mL}{l}(i+\omega\tau) 
\eta-\frac{\pi}{4}\right]}.
\label{eq:beforethetwist} 
\end{eqnarray} 
The two terms in the above sum will be handled separately.
Later, it will become clear, that the leading contribution to the integral
comes from the plus-sign term. Therefore, for the time being,
we ignore the term with the minus sign. 
Using the Cauchy theorem we can deform the contour of the integrals
such that its direction near the edge at $\eta =0$ is a steepest descent 
direction.  The contour is further deformed to follow 
steepest descent curves as illustrated in Fig.~\ref{fig:cutss}. 
Thus the contour 
consists of four segments: (a) from 0 to $i\infty/(1+ i/\omega \tau)$,  
(b) the part surrounding the cut, (c) an arc connecting these two 
segments at infinity, and (d) an arc connecting the end of the (b) path and  
the original end point at $(\omega\tau-i)\infty.$ 

It is straightforward to check that the contribution from
part (a) is of order $1/(\omega\tau)^{7/2}$. This contribution will
turn out to be negligible compared to the one coming
form segment (b). It is also clear that the contributions
of arcs (c) and (d) vanish, when their distance form the origin 
approaches infinity. Thus we focus our attention on segment (b).  

To evaluate the integral we exponentiate the pre-exponent 
factor in Eq.~(\ref{eq:beforethetwist}), 
and write the integral in the form
\begin{eqnarray}
K(m)\simeq\frac{e^{-i\frac{\pi}{4}}}{\sqrt{m L(2\pi l)^{3}(i+\omega\tau)^3}} 
\int e^{A(\eta)}d\eta, \nonumber 
\end{eqnarray} 
where
\begin{eqnarray}
A(\eta)=i \frac{mL}{l}(i+\omega\tau)\eta-\frac{3}{2}\ln(\eta-1). \nonumber
\end{eqnarray} 
The steepest decent contour is found in the usual way:
First, one takes the derivative of $A(\eta)$ with respect to $\eta$, 
\begin{eqnarray} 
A'(\eta)=i \frac{mL}{l}(i+\omega\tau)-3/2\frac{1}{\eta-1}, \nonumber
\end{eqnarray}
and find the saddle point $\eta^*$, satisfying  $A'(\eta^*)=0$.
The result is 
\begin{eqnarray}
\eta^{*}= u^{*}+iv^{*}=1-\frac{3l}{2mL(1-i\omega\tau)}, \nonumber
\end{eqnarray}
where $u^{*}$ and $v^{*}$ are real numbers.
Second, one substitutes
$\eta =(u-u^{*})+i(v-v^{*})$, where u and v are real, and solve for 
the curve which satisfies the condition
\begin{eqnarray}
\mbox{Im}(A(\eta))=\mbox{Im}(A(\eta^{*}))\nonumber  
\end{eqnarray} 
The formula for this curve is 
\begin{eqnarray}
\frac {v-\omega u}{u+\omega v -\frac{3}{2}}=\tan 
\frac{2}{3}(\omega u-v). \nonumber
\end{eqnarray}  
Rotating the axis as $x=v-\omega u$ and $y=u+\omega v$, 
one obtains 
\begin{eqnarray}
y=\frac{3}{2}-\frac{x}{\tan\frac{2}{3}x}. \nonumber 
\end{eqnarray} 
This exact form shows that the contour can be indeed deformed as 
shown in Fig.~13. It also provides the possibility of 
constructing the full asymptotic expansion of the integral. However,
in view of the approximations which already have been made,  
we are interested only in the leading term. 

Thus, taking the quadratic approximation for the action: 
$A(\eta) \simeq A(\eta^*) + \frac{1}{2} 
A''(\eta^*)(\eta-\eta^*)^2$, with
\begin{eqnarray}
A''(\eta)=\frac{3}{2}\frac{1}{(\eta-1)^{2}}, \nonumber
\end{eqnarray}
and evaluating the resulting Gaussian integral we arrive at
(\ref{eq:Kresult}). Notice that the result is proportional to 
$1/\omega \tau$. Thus, the edge contribution (which is of order
$1/(\omega\tau)^{7/2}$) can be indeed neglected. This calculation
also shows that $A''(\eta^*) \propto m^2$, and therefore
the small parameter of this saddle point 
approximation is $1/m$.

The contribution of the second term in the sum 
(\ref{eq:beforethetwist}), i.e the one with the minus sign, is calculated 
following along the same lines. However, it turns out that in this 
case the deformed contour does not pass through a saddle point, 
and the only contribution comes from the edge at $\eta =0$. It is, again, 
of  order $1/(\omega\tau)^{7/2}$,
and therefore can be neglected.

\end{multicols} 

\vskip -0.5 cm
\begin{references}
\vspace{-1.5 cm}
\bibitem {Heller84} E.~J.~Heller, Phys. Rev. Lett. {\bf 53}, 1515 (1984).  
\bibitem{Berry85} M. V. Berry, Proc. Roy.   
Soc. London {\bf A 400}, 229 (1985).  
\bibitem{Anderson58} P.~W.~Anderson, Phys. Rev. {\bf 109} 1492 (1958).  
\bibitem{Bergmann} G.~Bergmann, Phys. Rep. {\bf 107}, 1 (1984).  
\bibitem{advances} See for example   
U.~Sivan, F.~P.~Milliken, K.~Milkove, S.~Rishton, Y.~Lee, J.~M.~Hong, 
V.~Boegly, D.~Kern and M.~DeFranza, Europhys. Lett. {\bf 25}, 605 (1994);  
C.~M.~Marcus, S.~R.~Patel, A.~G.~Huibers, S.~M.~Cronenwett, M.~Switkes, 
I.~H.~Chan, R.~M.~Clarke, J.~A.~Folk, S.~F.~Godijn, K.~Campman, 
and A.~C.~Gossard,
Chaos Soliton Fract. {\bf 8}, 1261 (1997); 
D.~C.~Ralph, C.~T.~Black, and M.~Tinkham, Physica, {\bf 218V}, 258 (1996);
A.~Yacoby, R.~Schuster, and M.~ Heiblum, Phys. Rev. B {\bf 53},
9583 (1996); T.~M.~Fromhold, L.~Eaves, F.~W.~Sheard, 
T.~J.~Foster, M.~L.~Leadbeater and P.~C.~Main, Phys. Rev. Lett. {\bf 72}, 
2608 (1994); G.~Muller, G.~S.~Boebinger, H.~Mathur, L.~N.~Pfeiffer,
and K.~W.~West, Phys. Rev. Lett. {\bf 75}, 2875 (1995); 
D.~Goldhaber-Gordon, H.~Shtrikman, 
D.~Mahalu, D.~Abusch-Magder, U.~Meirav, and M.~A.~Kastner, 
Nature, {\bf 391}, 156 (1998).
\bibitem{Whitney99} R.~S.~Whitney, I.~V.~Lerner, and R.~A.~Smith,  
Wave Random Media {\bf 9}, 179 (1999).  
\bibitem{NLSM}  B.~A.~Muzykantskii and D.~E.~Khmelnitskii,    
JETP Lett. {\bf 62}, 76 (1995); A.~V.~Andreev, O.~Agam, B.~D. Simons,   
and B.~L. Altshuler, Phys. Rev. Lett. {\bf 76}, 3947 (1996).  
\bibitem{Aleiner} I.~L.~ Aleiner, and A.~I.~Larkin,  
Phys. Rev. B, {\bf 54}, 14423, (1996); Phys. Rev. E, {\bf 55}, R1243,   
(1997).  
\bibitem{Mehta91} M.~L.~Mehta, {\em Random Matrices and Statistical  
Theory of Energy Levels} (Academic Press, New York, 1991).  
\bibitem{roughb} F.~Borgonovi, G.~Casati, and B.~Li, 
Phys. Rev. Lett. {\bf 77}, 4744 (1996);
K.~M.~Frahm and D.~L.~Shepelyansky, Phys. Rev. Lett. {\bf 78}, 
1440 (1997); K.~M.~Frahm, Phys.~Rev.~B, {\bf 55}, R8626 (1997);
K.~M.~Frahm and D.~L.~Shepelyansky, Phys. Rev. Lett. {\bf 79}, 
1833 (1997).
\bibitem{Shepelyansky86} D.~L.~Shepelyansky, Phys. Rev. Lett. {\bf 56},  
677 (1986). 
\bibitem{Gutzwiller} M. C. Gutzwiller, {\it Chaos in Classical and  
Quantum Mechanics} (Springer, N.Y., 1990).  
\bibitem{Hannay84} J.~H.~Hannay and A.~M.~Ozorio de Almeida, J. Phys.
A {\bf 17}, 3429 (1984).
\bibitem{Smith98} R.~A.~Smith, I.~V.~Lerner, and B.~L.~Altshuler,
Phys. Rev. B {\bf 58}, 10343 (1998).
\bibitem{Altland93} A.~Altland and Y.~Gefen, Phys. Rev. Lett. 
{\bf 71}, 3339 (1993); Phys. Rev. B {\bf 51}, 
10671 (1995). 
\bibitem{Agam96d} O.~Agam and S.~Fishman, Phys. Rev. Lett. {\bf 76}, 
726 (1996); J. Phys. A: Math. Gen. {\bf 29}, 2013 (1996).
\bibitem{argaman} N.~Argaman, Y.~Imry, and U.~Smilansky, Phys. Rev. B {\bf 47}, 
4440 (1993). 
\bibitem{chalker} J.~T.~Chalker, I.~V.~Lerner, and R.~A.~Smith, Phys. 
Rev. Lett. {\bf 77}, 554 (1996); J. Math. Phys. {\bf 37}, 5061 (1996).
\bibitem{Abrikosov} A.~A.~Abrikosov, L.~P.~Gor'kov, and 
I.~E.~Dzyaloshinski, Methods of Quantum Field Theory 
 in Statistical Physics (Dover, New York, 1975).
\bibitem{Agam00} This relation is derived in O.~Agam  Phys. Rev. 
E {\bf 61}, 1285 (2000).
\bibitem{Berry72} M.~V.~Berry and K.E. Mount, Rep. Phys. {\bf 35},315 (1972).
\bibitem{Hikami81} S.~Hikami, Phys. Rev. B {\bf 24}, 2671 (1981).
\bibitem{Thouless74} D.~J.~Thouless, Phys. Rep.  {\bf 13} 93 (1974).
\bibitem{altshulershklovskii} B.~L.~Altshuler
and B.~I.~Shklovskii, JETP {\bf 64}, 127 (1986).
\bibitem{Kravtsov95} V.~E.~Kravtsov and I.~V.~Lerner, Phys. Rev. 
Lett. {\bf 74}, 2563 (1995).
\bibitem{Mandelbrot82} B.~Mandelbrot, {\em The Fractal Geometry in Nature}
(Freeman, San Francisco, 1982). 
\bibitem{Altshuler97} A similar situation appears in the Kepler billiard, 
see B.~L.~Altshuler and L.~S.~Levitov, Phys. Rep. {\bf 288}, 487 (1997). 
\bibitem{Agam94} O.~Agam and S.~Fishman, Phys. Rev. Lett.  
{\bf 73}, 806 (1994). 
\bibitem{Wilke00} J.~Wilke, A.~D.~Mirlin, D.~G.~Polyakov, F.~Evers, and
P.~Wolfle,  Phys. Rev. B {\bf 61}, 13774 (2000). 
\end{references}
\end{document}